\documentclass[aps,pra,showpacs,amsmath,amssymb,amsfonts,lengthcheck,twocolumn,longbibliography,superscriptaddress]{revtex4-2}

\usepackage{graphicx}
\usepackage{dcolumn}
\usepackage{bm}
\usepackage{braket}
\usepackage{enumitem}
\usepackage[utf8]{inputenc}
\usepackage{amsmath}
\usepackage{amssymb}
\usepackage{amsthm}
\usepackage{float}
\usepackage[colorlinks=true,citecolor=blue,linkcolor=blue,urlcolor=blue]{hyperref}

\newcommand{\tr}[1]{\mathrm{tr}\left\{#1\right\}}

\newcommand{\mbb}[1]{\mathbb{#1}}

\newcommand{\mrm}[1]{\mathrm{#1}}
\newcommand{\la}{\left\langle}
\newcommand{\ra}{\right\rangle}
\newcommand{\pd}{\partial}
\newcommand{\bla}{bla\\bla\\bla\\bla\\bla}

\newcommand{\revAB}[1]{{\color{cyan}#1}}

\begin{document}

\title{Quantum speed limit for observables from quantum asymmetry}

\author{Agung Budiyono}
\email{agungbymlati@gmail.com}
\affiliation{Research Center for Quantum Physics, National Research and Innovation Agency, South Tangerang 15314, Republic of Indonesia}
\affiliation{Department of Physics, University of Maryland, Baltimore County, Baltimore, MD 21250, USA}
\affiliation{Quantum Science Institute, University of Maryland, Baltimore County, Baltimore, MD 21250, USA}

\author{Michael Moody}
\affiliation{Department of Physics, University of Maryland, Baltimore County, Baltimore, MD 21250, USA} 
\affiliation{Quantum Science Institute, University of Maryland, Baltimore County, Baltimore, MD 21250, USA}

\author{Hadyan L. Prihadi}
\affiliation{Research Center for Quantum Physics, National Research and Innovation Agency, South Tangerang 15314, Republic of Indonesia} 

\author{Rafika Rahmawati}
\affiliation{Research Center for Quantum Physics, National Research and Innovation Agency, South Tangerang 15314, Republic of Indonesia} 

\author{Sebastian Deffner}
\affiliation{Department of Physics, University of Maryland, Baltimore County, Baltimore, MD 21250, USA} 
\affiliation{Quantum Science Institute, University of Maryland, Baltimore County, Baltimore, MD 21250, USA}
\affiliation{National Quantum Laboratory, College Park, MD 20740, USA}

\date{\today}

\begin{abstract} 
Quantum asymmetry and coherence are genuinely quantum resources that are essential to realize quantum advantage in information technologies. However, all quantum processes are fundamentally constrained by quantum speed limits, which raises the question on the corresponding bounds on the rate of consumption of asymmetry and coherence. In the present work, we derive a formulation of the quantum speed limit for observables in terms of the trace-norm asymmetry of the time-dependent quantum state relative to the observable. This quantum speed limit can be directly observed in experiment through weak value measurement and provides a lower bound to the quantum Fisher information about the parameter conjugate to the observable. It can be further related to quantum coherence relative to the eigenbasis of the observable. We obtain a complementary relation for the speed of three mutually unbiased observables for a single qubit. As an application, we derive a notion of a \emph{quantum thermodynamic speed limit}. 
\end{abstract} 

\maketitle       

\section{Introduction}

The quantum speed limit is a fundamental restriction on the maximum speed with which a quantum system can evolve \cite{Frey2016QINP,Deffner2017JPA}. Thus, it sets a crucial constraint in different information-processing protocols leveraging quantum dynamics such as quantum computation and communication \cite{Lloyd2000Nature,Ashhab2012PRA,Deffner2020PRR}, quantum control \cite{Canvea2009PRL,Campbell2017PRL,Aifer2022NJP}, quantum metrology \cite{Giovannetti2011NPhot,Campbell2018QST}, quantum thermodynamics \cite{Deffner2010PRL,Mukhopadhyay2018PRE,Funo2019NJP,Deffner2025QST}, and recently in the speed of quantum resources \cite{Mohan2022NJP,Campaioli2022NJP,Allan2021Quantum,Pratapsi2025QST,Marvian2016PRA_2}.

The first example of a genuine quantum speed limit was derived by Mandelstam and Tamm as an operational and a rigorous reformulation of the Heisenberg uncertainty relation for energy and time \cite{Mandelstam1945}. Since then, extensive efforts have been made to study its rich aspects from various angles and along different directions \cite{Bhattacharyya1983JPA,Fleming1973,Anandan1990PRL,Pati1991PLA,Uhlmann1992PLA,Vaidman1992AJP,Uffink1993AJP,Margolus1998PD,Kupferman2008PRA,Levitin2009PRL,Jones2010PRA,Taddei2013PRL,Campo2013PRL,Deffner2013PRL,Pires2016PRX,Campaioli2018PRL,OConnor2021PRA,Poggi2021PRXQ,Hamazaki2022PRXQ,Deffner2022EPL}. The usual approach to define quantum speed is based on the state distinguishability via some notion of distance in state space. Note, however, that there are pairs of quantum states which are perfectly distinguishable in the state space, but their expectation values relative to a physically relevant observable are equal or very close to each other. This observation has led to recent efforts to study quantum speed and its limit based on using the speed of the quantum expectation value of an observable in a state \cite{Mohan2022PRA,Pintos2022PRX,Shrimali2024PRA}. A related notion of speed of operator growth is discussed in Refs. \cite{H_rnedal_2022,Carabba:2022aa}. In the present work, we discuss the relation between the speed of expectation value of an observable and quantumness arising from the noncommutativity between the observable and the state.  

Consider a quantum system with a time-dependent state $\varrho(t)$ \revAB{acting} on a Hilbert space $\mathcal{H}$. Then the quantum ``speed'' associated with an observable $K$ is given by the rate of change of its expectation value \cite{Mohan2022PRA,Pintos2022PRX,Shrimali2024PRA},
\begin{eqnarray}
v_K\equiv\frac{1}{2}\left|\pd_t\la K\ra\right|,
\label{quantum speed of observable}
\end{eqnarray}
where for any operator $O$ the angular brackets denote an average over the time-dependent state, $\la O\ra =\tr{O\,\varrho(t)}$. In the definition \eqref{quantum speed of observable}, we have included a factor $1/2$ for later convenience. Interestingly, this notion of speed is also the starting point for deriving the Mandelstam-Tamm bound \cite{Mandelstam1945}, where for pure $\varrho(t)$ the observable $K$ is chosen as the projector onto  $\varrho(t)$.

For general $K$ and unitary dynamics $U(t)$, with generator $H(t)=-i\, \dot{U}(t)^{\dagger}U(t)$, it can be shown that \cite{Mohan2022PRA,Mohan2022PRA,Shrimali2024PRA},
\begin{equation}
v_K\leq\Delta_K\,\Delta_H\,,  
\label{quantum speed of observable under unitary dynamics}
\end{equation}
where $\Delta_O^2=\la O^2\ra-\la O\ra^2$ is the variance of observable $O$. As usual, we are working in units where $\hbar=1$. This result has recently been rigorously generalized to arbitrary quantum dynamics \cite{Pintos2022PRX,Mohan2022PRA,Shrimali2024PRA}. 

For our present purposes, we note that the instantaneous quantum speed \eqref{quantum speed of observable} under unitary dynamics is governed by the pairwise quantum noncommutativity relations between the generator of the dynamics, the probing observable, and the state. Noncommutativity crucially underlies quantum coherence and asymmetry \cite{Marvian2016PRA,Gour2008NJP,Vaccaro2008PRA,Budiyono2023PRA_2,Budiyono2023JPA,Streltsov2017RMP,Baumgratz2014PRL,Girolami2014PRL,Budiyono2023PRA,Budiyono2024JPA}, and general quantum correlation including quantum entanglement between two or more parties \cite{Adesso2016JPA,Budiyono2023JPAab,Horodecki2009RMP,Budiyono2025PRA}, all of which are necessary resources for quantum information protocols. It is therefore natural to ask if these quantumness as resource can be used to fully characterize $v_K$.

Yet, the usual formulation of upper bounds on $v_K$, such as Eq.~\eqref{quantum speed of observable under unitary dynamics}, obscures this issue. While the instantaneous speed must be vanishing when the probing observable $K$ commutes with the state $\varrho(t)$, the upper bound, which depends on the variance of $K$, can still be finite \footnote{This is the case when the state $\varrho(t)$ is a classical statistical mixture of the eigenbasis $\{\ket{k}\}$ of $K$, i.e., $\varrho(t)=\sum_kp_k\ket{k}\bra{k}$, for some set of normalized probabilities $\{p_k\}$, $p_k\ge 0$, $\sum_kp_k=1$.} Therefore, the present work is dedicated to deriving an upper bound, i.e., a version of the quantum speed limit that is explicitly determined by the noncommutativity between the probing observable $K$ and the state, $\varrho(t)$. Through such a formulation, we can then directly relate the quantum speed limit for an observable with quantum asymmetry and coherence carried by a quantum state relative to the observable, $K$. 

\section{Speed and quantum asymmetry \label{Quantum speed of observable and asymmetry}}

Consider a quantum system with a finite-dimensional Hilbert space $\mathcal{H}$ undergoing unitary dynamics. For such scenarios, if infinite resources are available, also the quantum speed can be infinite. In particular, if the spectrum of the Hamiltonian is unbounded, i.e., $\|H(t)\|_{\infty}=\infty$, the right hand side of Eq.~\eqref{quantum speed of observable under unitary dynamics} can diverge. We denote the operator norm by $\|H(t)\|_{\infty}$, which is identical to the Schatten-$\infty$ norm of $H(t)$. Since this case is not particularly interesting, we restrict ourselves to a set of unitaries with a bounded Hamiltonian generator. 

For simplicity and without loss of generality we then consider only Hamiltonians, for which $\|H(t)\|_{\infty}=1$. To ease the notation, we denote the corresponding quantum speed limit as
\begin{equation}
\label{eq:QSL}
v_\mrm{QSL}\equiv \sup_{\|H(t)\|_{\infty}=1}\left\{v_K\right\}\,,
\end{equation}
which is the maximal quantum speed a system with bounded Hamiltonian can achieve.

Inspecting this quantum speed limit \eqref{eq:QSL}, we note that the state $\varrho(t)$ can be understood as a quantum object containing a useful quantum resource. This resource can be harnessed using a probing observable $K$ via a protocol or strategy encoded in the unitary evolution $U(t)$ generated by $H(t)$. We then optimize the instantaneous speed or the power of harnessing or injecting the resource content over all the unitary protocols. Interestingly, the quantum ergotropy is defined in a similar spirit as the maximum amount of work that can be extracted using all allowed unitaries \cite{Allahverdyan2004EPL}. Therefore, $v_\mrm{QSL}$ can be interpreted as the optimal power of harnessing or injecting a certain useful resource captured by the expectation value of $K$ in the state $\varrho(t)$ using any allowed unitary protocols, relative to the worst cost of implementing the unitary protocol.

To derive an upper bound on $v_\mrm{QSL}$ we first express the instantaneous speed as 
\begin{equation}
v_K=\frac{1}{2}\left|\tr{[H(t),\varrho(t)] K}\right|=\frac{1}{2}\left|\tr{H(t),[\varrho(t),K]}\right|\,.
\end{equation}
The latter can be further bounded by H\"older's inequality \cite{Baumgartner2011} which reads $\left|\tr{A\,B^\dagger}\right|\leq\|A\|_p\|B\|_q$ with $1/p+1/q=1$, $(p,q)\in[1,\infty)$. Here $\|O\|_p\equiv\tr{|O|^p}^{1/p}$, $|O|=\sqrt{OO^{\dagger}}$, is the Schatten-$p$ norm of $O$, which can be expressed in terms of the singular values. For instance, the Schatten-1 norm of $O$, i.e., the trace norm can be written as $\|O\|_1=\sum_\nu \sigma_\nu$, where $\sigma_{\nu}$ are the singular values of $O$. Now choosing $A=H(t)$ with $p=\infty$ and $B=[\varrho(t),K]$ with $q=1$, we immediately have
\begin{equation}
v_K \leq \frac{1}{2} \|H(t)\|_\infty\,\|[\varrho(t),K]\|_1\,.
\end{equation}
Thus, for bounded generators with $\|H(t)\|_\infty=1$, we obtain the quantum speed limit as the trace-norm noncommutativity between the state and the probing observable
\begin{equation}
v_\mrm{QSL}\leq\frac{1}{2}\,\|[\varrho(t),K]\|_1. 
\label{eq:ourQSL}
\end{equation}
Interestingly, as shown in App.~\ref{Proof of equality in eq. 5} for a single qubit, the inequality becomes an equality where the supremum is attained when the generator $H(t)$ of the unitary dynamics at time $t$ has eigenvalues $\{-1,1\}$.

Equation~\eqref{eq:ourQSL} is our first main result. The upper bound, that is trace-norm noncommutativity $\|[\varrho(t),K]\|_1/2$, is a bonafide measure of the asymmetry, called trace-norm asymmetry, of the state $\varrho(t)$ relative to the group of unitary translations $\{U_{\theta}=e^{-iK\theta}\}_{\theta}$, $\theta\in\mathbb{R}$, generated by the Hermitian operator $K$ \cite{Marvian2016PRA}. Hence, we observe that the instantaneous speed of the expectation value of the observable $K$ in the state $\varrho(t)$ under unitary dynamics with Hermitian generators having an operator norm unity, is upper bounded by the trace-norm asymmetry in the state $\varrho(t)$ relative to the group of unitary translations generated by $K$.

Alternatively, Eq.~\eqref{eq:ourQSL} can also be understood as the maximum speed of harnessing a feature captured by the expectation value of the probing observable $K$ in the state $\varrho(t)$ using any unitary protocol, relative to the worst cost of implementing the unitary. This quantum speed is upper bounded by the quantum asymmetry in the state relative to the group of translation unitaries generated by the probing observable $K$.

It is also interesting to point out that related results have appeared in the literature. In particular, Ref. \cite{Marvian2016PRA} derived a quantum speed limit for unitary dynamics in terms of the trace-norm asymmetry $\|[\varrho(t),H]\|_1/2$ relative to the generator of the unitary, $H$. Similar results were obtained in Ref. \cite{Mondal2016PLA}, where the asymmetry is quantified by the Wigner-Yanese skew information \cite{Wigner1963PNAS} and quantum Fisher information \cite{Sekiguchi:2024aa}. By contrast, in our result \eqref{eq:ourQSL} the quantum asymmetry is determined relative to a specific observable $K$, which is \emph{not} the generator of and thus decoupled from the dynamics. In the following, we will see that our formulation of the quantum speed limit is, thus, particularly useful in quantum thermodynamic considerations.

\subsection{Quantum speed from weak measurements}

We begin by discussing the operational, statistical, and information theoretical meaning of our formulation of the quantum speed limit for observables \eqref{eq:ourQSL}. In previous work \cite{Budiyono2023PRA_2,Budiyono2023JPA}, it was shown that the trace-norm asymmetry in the state $\varrho(t)$ relative to the group of translation unitaries generated by $K$ can be expressed variationally as 
\begin{equation}
\frac{1}{2}\,\|[K,\varrho(t)]\|_1=\sup_{\mathbb{B}_o(\mathcal{H})} \left\{\sum_x\left|{\mathfrak{I}}\{K_{\rm w}(x|\varrho(t))\}\right|\,{\rm Pr}(x|\varrho(t))\right\}, 
\label{trace norm asymmetry in terms of optimal imaginary weak value}
\end{equation}
where we have introduced the weak value of $K$ with the preselected state $\varrho(t)$ and postselected state $\ket{x}$ \cite{Aharonov1988PRL,Wiseman2002PRA},
\begin{equation} 
K_{\rm w}(x|\varrho(t)):=\frac{\braket{x|K\varrho(t)|x}}{\braket{x|\varrho(t)|x}}\,.
\end{equation}
Moreover, ${\rm Pr}(x|\varrho(t))=\tr{\Pi_x\,\varrho(t)}$ is the probability to obtain the outcome $x$ in a measurement described by the rank-1 orthogonal PVM (projection-valued measure) $\{\Pi_x:=\ket{x}\bra{x}\}$. The supremum is taken over all $\{\ket{x}\}\in \mathbb{B}_o(\mathcal{H})$, which is the set of all orthonormal bases of $\mathcal{H}$.

Therefore, we also have
\begin{equation}
v_\mrm{QSL}\leq \sup_{\mathbb{B}_o(\mathcal{H})} \left\{\sum_x \left|{\mathfrak{I}}\{K_{\rm w}(x|\varrho(t))\}\right|\,{\rm Pr}(x|\varrho(t))\right\}\,,
\label{the optimal speed of expectation value is upper bounded by the average imaginary weak value} 
\end{equation}
and hence the quantum speed is experimentally accessible. In fact, the weak value $K_{\rm w}(x|\varrho(t))$ can be directly observed in experiments, either using a weak measurement followed by a strong projection measurement \cite{Aharonov1988PRL,Wiseman2002PRA,Lundeen2005PLA,Jozsa2007PRA}, or using other methods without using weak measurements \cite{Johansen2007PLA,Vallone2016PRL,Cohen2018PRA,Wagner2024QST,Chiribella2024PRR}.  

Hence, both sides of Eq. (\ref{the optimal speed of expectation value is upper bounded by the average imaginary weak value}) can be in principle inferred directly in experiment, whilst involving optimization, respectively, over all unitaries on and all orthonormal bases of the Hilbert space. Such optimizations can be implemented using the presently available NISQ hardware by employing hybrid quantum-classical variational quantum algorithm \cite{Cerezo2021NRP}. Recently, the imaginary part of the weak value has been argued to indicate quantum contextuality via weak measurement with postselection \cite{Kunjwal2019PRA} and a notion of epistemic restriction underlying quantum uncertainty \cite{Budiyono:2021aa}. In this sense, Eq. (\ref{the optimal speed of expectation value is upper bounded by the average imaginary weak value}) thus shows that the quantum contextuality and epistemic restriction are necessary for a nonvanishing quantum speed of expectation value, and conversely, a nonvanishing quantum speed of expectation value is sufficient to indicate quantum contextuality and epistemic restriction. The imaginary part of the weak value $K_{\rm w}(x|\varrho(t))$ can also be interpreted as a disturbance due to the weak measurement of $K$ \cite{Aharonov2005PRA,Dressel2012PRA}. Hence, according to Eq.~\eqref{the optimal speed of expectation value is upper bounded by the average imaginary weak value}, the sensitivity of the state due to weak measurement of $K$ gives an upper bound to the speed of the expectation value of $K$ under unitary dynamics. 

\subsection{Quantum metrological consequences}

Our quantum speed limit for observables \eqref{eq:ourQSL} can also be directly related to the precision with which the values of a real parameter conjugate to the observable $K$ can be estimated in quantum measurements. In fact, in Ref.~\cite{Budiyono2023PRA_2} it was shown that the trace-norm asymmetry is upper bounded by 
\begin{equation}
\|[\varrho(t),K]\|_1\le\sqrt{\mathcal{F}_{\theta}(\varrho(t),K)}\,,
\label{trace norm asymmetry is upper bounded by the quantum Fisher information}
\end{equation}
where $\mathcal{F}_{\theta}(\varrho(t),K)$ is the quantum Fisher information \cite{Braunstein1994PRL} about a parameter $\theta$ that is imprinted onto the quantum state $\varrho(t)$ by a unitary translation generated by $K$, $\varrho_{\theta}(t)=\exp{(-iK\theta)}\varrho(t)\exp{(iK\theta)}$. Note that the inequality in Eq. (\ref{trace norm asymmetry is upper bounded by the quantum Fisher information}) becomes an equality for pure states and for all states of a single qubit.

Accordingly, our quantum speed limit \eqref{eq:ourQSL} can also be written as
\begin{equation}
\label{quantum speed of expectation value is upper bounded by the quantum Fisher information}
v_\mrm{QSL}\leq \frac{1}{2}\,\sqrt{\mathcal{F}_{\theta}(\varrho(t),K)}\,.
\end{equation}
Thus, we conclude that the quantum speed of the observable $K$ sets a lower bound on the amount of information that can be extracted about a parameter $\theta$ conjugate to $K$.

It has been well established in the literature \cite{Taddei2013PRL} that $\sqrt{\mathcal{F}_{\theta}(\varrho(t),K)}/2$ is the natural, geometric quantum speed on the $\theta$-parameterized manifold of states. In fact, this is the maximum quantum speed any dynamics can achieve \cite{OConnor2021PRA}. We have now shown that the maximum instantaneous speed of the observable $K$ for any unitary dynamics with bounded generators is always lower than the speed of state evolution induced by the unitary generated by $K$. 

However, while our $v_\mrm{QSL}$ is the speed associated with the unitary dynamics, the upper bound in \eqref{quantum speed of expectation value is upper bounded by the quantum Fisher information} is the speed in the parameter $\theta$ conjugate to $K$ induced by the unitary translation $\exp{(-iK\theta)}$. In fact, Eq.~\eqref{quantum speed of expectation value is upper bounded by the quantum Fisher information} can be read as a relation between the speed of quantum evolution and the rate with which a parameter $\theta$ can be measured.

\subsection{Quantum speed limit and quantum fluctuations}

Our quantum speed limit \eqref{eq:ourQSL} can also be directly related to established formulations for observables \eqref{quantum speed of observable under unitary dynamics}. In fact, we have
\begin{equation}
\frac{1}{2}\|[K,\varrho(t)]\|_1\leq \Delta_K, 
\label{trace-norm asymmetry is upper bounded by the quantum standard deviation}
\end{equation}
where the inequality again becomes an equality for pure states \cite{Marvian2016PRA}. When $K$ and $\varrho(t)$ commute, the left-hand side vanishes while the right-hand side may remain finite. This fact suggests that the trace-nom asymmetry $\|[K,\varrho(t)]\|/2$ can be understood as the genuine quantum part of the fluctuations of $K$, while $\Delta_K$ quantifies the total (quantum and classical) fluctuations of $K$ relative to $\varrho(t)$ \cite{Luo2005PRA,Korzekwa2014PRA,Hall2023PRA,Budiyono2024JPA}. 

Noting this, Eq.~\eqref{eq:ourQSL} can be interpreted that the maximum speed of the quantum expectation value of an observable $K$ in the state $\varrho(t)$, under any unitary dynamics with a Hermitian generator having operator norm unity, is upper bounded by the genuine quantum fluctuations of $K$ in $\varrho(t)$. By contrast, in classical stochastic mechanics, the rate of change of the average of a random variable is not fundamentally restricted by the fluctuations of the variable.

\subsection{Quantum speed limit time}

So far we have focused on the quantum speed limit, i.e., on upper bounds on the rate with which an expectation value can change. Associated with any maximal rate is always a minimal time, which in the present case corresponds to the minimal time that is required to observe a predetermined change in the expectation value of $K$. 

To this end, we now need to specify the parametrization of the time-dependent Hamiltonian generating the unitary dynamics. Thus, we write $H[\gamma(t)]$, where $\gamma(t)$ is a time-dependent protocol. The corresponding time-evolution then becomes
\begin{equation}
U[\gamma(t)]=\mathcal{T}_>\exp{\left(-i\int^t \mrm{d}t'\,H[\gamma(t')]\right)}\,,
\end{equation}
where $\mathcal{T}_>$ is the time ordering operator.

Integrating Eq.~\eqref{quantum speed of observable} and using Eqs.~\eqref{eq:QSL} and ~\eqref{eq:ourQSL} we obtain that for a change in expectation value $\Delta K_\tau=|\la K\ra_\tau-\la K\ra_0|$ in time $\tau$ the quantum system needs at least
\begin{equation}
\tau\geq\frac{\Delta K_\tau}{\overline{\|[\varrho[\gamma],K]\|_1}}\equiv \tau_\mrm{QSL}\,,
\label{minimum time for state transformation using unitary dynamics in terms of asymmetry}
\end{equation}
where $\overline{f[\gamma]}=1/\tau\,\int_{0}^{\tau}{\rm d}t\,f[\gamma(t)]$ is the time average of the function $f(t)$ along the trajectory $\gamma(t)$.  

In other words, given a unitary dynamics $U[\gamma(t)]$ with Hermitian generator $H[\gamma(t)]$  the minimum time needed to transform an initial state $\varrho(0)$ to a final state $\varrho(\tau)$ inducing a difference between the initial and final expectation value of the probing observable $\Delta K_\tau$, is inversely proportional to the time-average trace-norm asymmetry $\overline{\|[\varrho[\gamma],K]\|_1/2}$ of the state $\varrho(t)$ relative to the translation unitaries generated by $K$ along the trajectory. In particular, the larger the time-average trace-norm asymmetry generated along the trajectory, the smaller is the minimum time needed to get a targeted change of expectation value of the observable using the unitary transformation.

The obvious question then is whether there is an optimal choice for the parameterization $\gamma(t)$. Indeed, this simply becomes an optimal control problem for $\gamma(t)$, in the sense that the optimal $\gamma(t)$ is the protocol that minimizes the lower bound in Eq.~\eqref{minimum time for state transformation using unitary dynamics in terms of asymmetry}. Since the numerator in the lower bound does not depend on the trajectory, the optimal control is attained by $\gamma(t)$ that maximizes the trace-norm asymmetry.

Moreover, it is worth emphasizing that Eq.~\eqref{minimum time for state transformation using unitary dynamics in terms of asymmetry} holds for any choice of $K$. In particular, also for $K$ that commute with the quantum state $\varrho(t)$. Hence, one can choose $K$ which solves the variational equation $\|\varrho(\tau)-\varrho(0)\|_1/2=\sup_{0\leq K\le\mathbb{I}}\tr{K(\varrho(\tau)-\varrho(0))}$, in which case Eq.~\eqref{minimum time for state transformation using unitary dynamics in terms of asymmetry} becomes the minimal time required for a quantum state $\varrho(0)$ to evolve to a another state $\varrho(\tau)$.

\section{Coherence and complementarity}
\label{Quantum speed limit of expectation value associated with coherence and complementarity relations}

Our quantum speed limit for observables \eqref{eq:ourQSL} can also be directly related to measures of coherence. In this context, recall that the asymmetry in the state $\varrho(t)$ relative to the group of unitary translations generated by a Hermitian observable $K$ has been dubbed \emph{unspeakable coherence} relative to the orthonormal eigenbasis $\{\ket{k}\}$ of $K$ \cite{Budiyono2023JPA,Marvian2016PRA,Budiyono2023PRA_2}.

Thus, it is interesting to analyze $v_\mrm{QSL}$ \eqref{eq:ourQSL} for specific choices of $\{\ket{k}\}$. To this end, we now consider bounded $K$, where again without loss of generality we can set $\|K\|_\infty=1$. In such cases, it was shown in Ref.~\cite{Budiyono2024PRA} that
\begin{equation}
\sup_{K}\left\{\frac{\|[\varrho(t),K]\|}{2\,\|K\|_{\infty}}\right\}\le C_{\rm KD}^{\rm NRe}(\varrho(t);\{\ket{k}\}), 
\label{normalized trace-norm asymmetry is upper bounded by quantum coherence}
\end{equation}
where $C_{\rm KD}^{\rm NRe}(\varrho(t);\{\ket{k}\})$ is a quantifier of coherence in the state $\varrho(t)$ relative to the orthonormal basis $\{\ket{k}\}$ defined as \cite{Budiyono2023PRA}
\begin{equation}
C_{\rm KD}^{\rm NRe}(\varrho(t);\{\ket{k}\}):=\sum_k\sup_{\mathbb{B}_{\rm o}(\mathcal{H})}\sum_c|{\rm Im}\{{\rm Pr}_{\rm KD}(k,c|\varrho(t))\}|,
\label{KD-nonreality coherence}
\end{equation}
and ${\rm Pr}_{\rm KD}(k,c|\varrho(t)):=\braket{c|k}\braket{k|\varrho(t)|c}$ is the Kirkwood-Dirac (KD) quasiprobability associated with the state $\varrho(t)$ relative to two orthonormal bases $\{\ket{k}\}$ and $\{\ket{c}\}$ of $\mathcal{H}$ \cite{Kirkwood1933PR,Dirac1945RMP}.

Accordingly, we also have that
\begin{equation}
\label{the speed of expectation value of normalized observable is upper bounded by the KD-nonreality coherence} 
v_\mrm{QSL}\leq C_{\rm KD}^{\rm NRe}(\varrho(t);\{\ket{k}\})\,.
\end{equation}
Thus, the quantum speed limit for observables is upper bounded by the coherence encoded in $\varrho(t)$ relative to the orthonormal eigenbasis of the observable. Interestingly, for single qubits the inequality in Eq.~\eqref{the speed of expectation value of normalized observable is upper bounded by the KD-nonreality coherence} becomes an equality when the two eigenvalues $\{k_+,k_-\}$ of $K$ satisfies $k_+=-k_-$~\cite{Budiyono2024PRA}.

Next, as shown in Ref.~\cite{Budiyono2023PRA}, the KD-nonreality coherence \eqref{KD-nonreality coherence} is upper bounded as 
\begin{eqnarray}
C_{\rm KD}^{\rm NRe}(\varrho(t);\{\ket{k}\})\le C_{l_1}(\varrho(t);\{\ket{k}\}),
\label{KD-nonreality coherence is upper bounded by the trace-norm coherence}
\end{eqnarray}
where, $C_{l_1}(\varrho(t);\{\ket{k}\})$ is the $l_1$-norm coherence in the state $\varrho(t)$ relative to the orthonormal basis $\{\ket{k}\}$ which is defined as $C_{l_1}(\varrho(t);\{\ket{k}\}):=\sum_{k\neq k'}|\braket{k|\varrho(t)|k'}|$ \cite{Baumgratz2014PRL}. Moreover, for any state of a single qubit, the inequality \eqref{KD-nonreality coherence is upper bounded by the trace-norm coherence} again becomes an equality.

\subsection{Example: quantum speed of single qubits}

Since we have mentioned several times above that the inequalities become tight for single qubits, we pause to illustrate and demonstrate our findings for this scenario. To this end, consider two unit vectors $\vec{n}_K\in \mbb{R}^3$ and $\vec{n}_U \in \mbb{R}^3$, and the Pauli vector $\vec{\sigma}=(\sigma_x,\sigma_y,\sigma_z)$. In this case, the quantum state is written in Bloch representation as $\varrho(t)=\frac{1}{2}(\mathbb{I}+\vec{r}_{\rm S}\cdot\vec{\sigma})$, and we choose as a probing observable $K=\vec{n}_K\cdot\vec{\sigma}$. Note that it is easy to see that the eigenvalues of $K$ are $\{1,-1\}$, and hence Eqs.~\eqref{the speed of expectation value of normalized observable is upper bounded by the KD-nonreality coherence} and \eqref{KD-nonreality coherence is upper bounded by the trace-norm coherence} become equalities.

Thus, we can also find a simple relation for the minimum time \eqref{minimum time for state transformation using unitary dynamics in terms of asymmetry} for the quantum system to achieve a change in expectation value. We have,
\begin{equation}
\label{minimum time in terms of time-average coherence}
\tau\ge\frac{|\tr{(\vec{n}_K\cdot\vec{\sigma})\varrho(\tau)}-\tr{(\vec{n}_K\cdot\vec{\sigma})\varrho(0)}|/2}{\overline{C_{l_1}(\varrho[\gamma];\{\ket{k}\})}}
\equiv \tau_{\rm min}\,,
\end{equation}
which follows from using the equality of Eqs.~\eqref{the speed of expectation value of normalized observable is upper bounded by the KD-nonreality coherence} and \eqref{KD-nonreality coherence is upper bounded by the trace-norm coherence} in the expression \eqref{minimum time for state transformation using unitary dynamics in terms of asymmetry}. As mentioned above, this inequality \eqref{minimum time in terms of time-average coherence} becomes time, for the optimal protocol $\gamma(t)$ determined from solving the corresponding optimal control problem. 

\begin{figure}
  \centering
  \includegraphics[width=.48\textwidth]{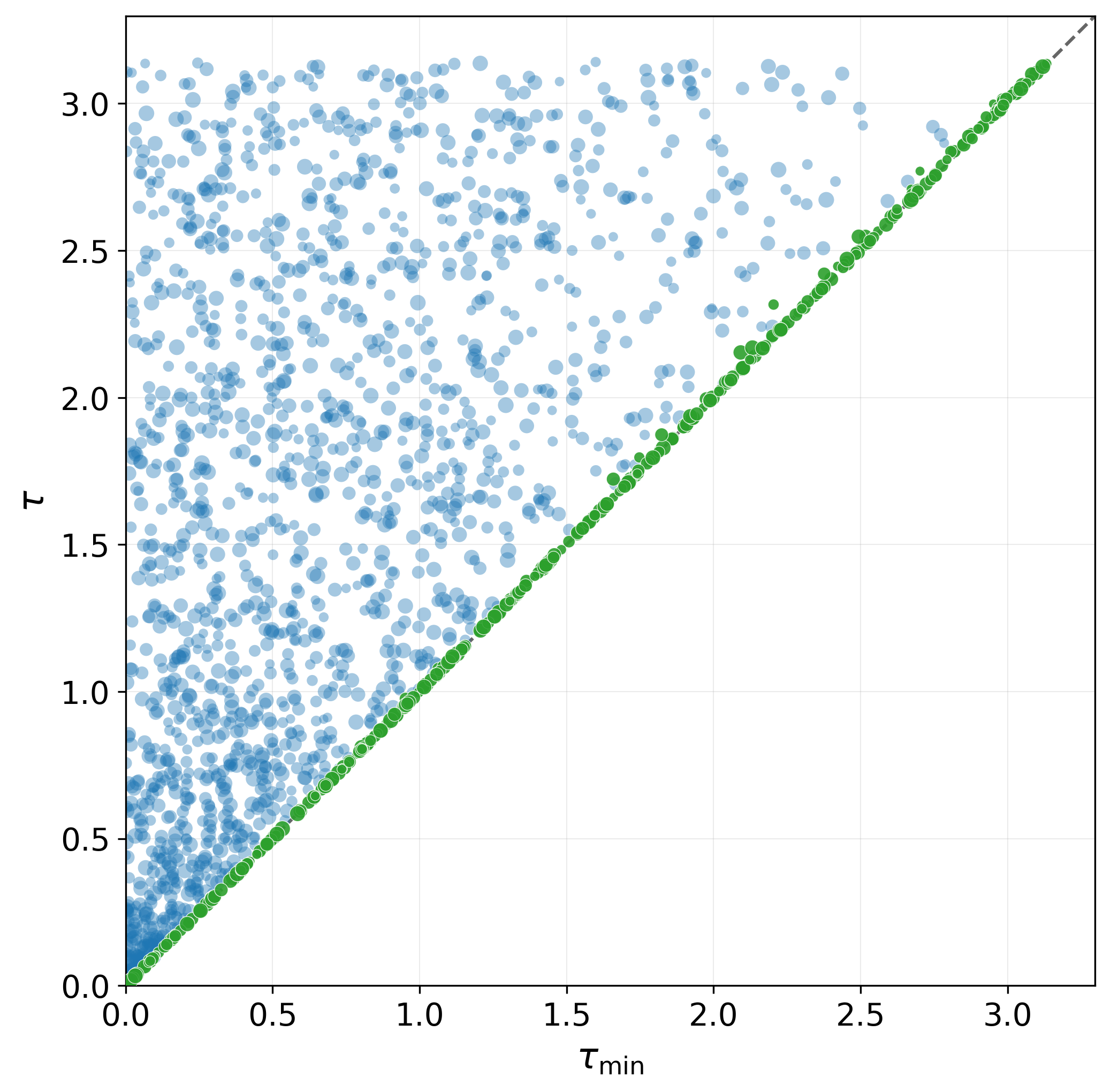}
  \caption{Realizations of the a quantum process (blue markers), with random initial states and unitary maps. The size of the marker represents the purity of the initial states. Green markers depict realizations that are optimal.}
  \label{fig:qsl_scatter}
\end{figure}

This is illustrated for a numerical solution of the problem in Fig.~\ref{fig:qsl_scatter}. To this end, we randomly chose initial states and unitary maps. Specifically, for each trial  we randomly sampled $\vec{n}_K$, $\vec{n}_U(0)$ and $\vec{r}_S(0)$, solve the von-Neumann equation for sufficiently short interval of time, and vary the control unitary by changing randomly $\vec{n}_U(t)$ for each time step. We computed the difference between the expectation value of $\vec{n}_K\cdot\vec{\sigma}$ and the time-average $l_1$-norm coherence relative to the eigenbasis of $\vec{n}_K\cdot\vec{\sigma}$ to obtain $\tau_{\rm min}$ \eqref{minimum time in terms of time-average coherence}. Observe, that all the trials satisfy the inequality. The green markers depict realizations that saturate the lower bound is obtained by randomly choosing $\vec{n}_U(t)$ that are orthogonal to the plane defined by $\vec n_{K}$ and $\vec r_{\mathrm S}(t)$ for all $t \in [0,\tau],$ in case of which the maximum speed is attained.

\subsection{Complementarity bounds for qubits}

The equalities in \eqref{the speed of expectation value of normalized observable is upper bounded by the KD-nonreality coherence} and \eqref{KD-nonreality coherence is upper bounded by the trace-norm coherence} for a single qubit further implies the following complementarity relations for the quantum speed of three observables whose eigenbases comprise a set of mutually unbiased bases (MUB) of the two-dimensional Hilbert space $\mathcal{C}^2$ \cite{Ivonovic1981JPA,Wootters1986FP,Wootters1989AP}. First, consider three mutually complementary Pauli operators, e.g., $\sigma_x$, $\sigma_y$ and $\sigma_z$. Note that the eigenbases of $\sigma_x$, $\sigma_y$ and $\sigma_z$, respectively, $\mathbb{X}=\{\ket{x_+},\ket{x_-}\}$, $\mathbb{Y}=\{\ket{y_+},\ket{y_-}\}$ and $\mathbb{Z}=\{\ket{z_+},\ket{z_-}\}$, where $\ket{x_{\pm}}=\frac{1}{\sqrt{2}}(\ket{z_+}\pm\ket{z_-})$ and $\ket{y_\pm}=\frac{1}{\sqrt{2}}(\ket{z_+}\pm i\ket{z_-})$, are mutually unbiased with each other. Moreover, in this case, the $l_1$-norm coherence in the state $\varrho(t)$ relative to these MUB bases satisfy the following complementarity relation \cite{Cheng2015PRA},
\begin{equation}
C_{l_1}(\varrho(t);\mathbb{X})^2+C_{l_1}(\varrho(t);\mathbb{Y})^2+C_{l_1}(\varrho(t);\mathbb{Z})^2=2|\vec{r}_{\rm S}|^2. 
\label{Cheng2015PRA for a single qubit}
\end{equation}
Thus, we can also write,
\begin{equation}
(v_\mrm{QSL}^\mbb{X})^2+(v_\mrm{QSL}^\mbb{Y})^2+(v_\mrm{QSL}^\mbb{Z})^2=2\left[2P(t)-1\right]\,,
\end{equation}
where $P(t)=\tr{\varrho(t)^2}=(1+|\vec{r}_{\rm S}|^2)/2$ is the state purity.

As a simpler but weaker form of a complementarity relation for our quantum speed limit, we note that as shown in Ref.~\cite{Mondal2017PRA}, the $l_1$-norm coherences in the state relative to the orthonormal bases $\mathbb{X}$, $\mathbb{Y}$, and $\mathbb{Z}$ satisfy a complementarity relation which is independent of the state purity,
\begin{equation}
C_{l_1}(\varrho(t);\mathbb{X})+C_{l_1}(\varrho(t);\mathbb{Y})+C_{l_1}(\varrho(t);\mathbb{Z})\le \sqrt{6}, 
\label{Mondal complementarity relation for ltrace-norm coherence} 
\end{equation}
and the inequality becomes an equality for a specific pure state of the form $\varrho(t)=\frac{1}{2}[\mathbb{I}+\frac{1}{\sqrt{3}}(\sigma_x+\sigma_y+\sigma_z)]$. Thus, we obtain
\begin{equation}
v_\mrm{QSL}^\mbb{X}+v_\mrm{QSL}^\mbb{Y}+v_\mrm{QSL}^\mbb{Z}\le \sqrt{6}. 
\label{complementarity relation for the speed limit of expectation value for a single qubit}
\end{equation}
The above results in particular shows that for any state $\varrho(t)$ the maximum speed of the expectation value of $\sigma_x$, $\sigma_y$ and $\sigma_z$ cannot all reach their speed limit.

\section{Relative entropy of athermality}
\label{Application: Quantum speed limit of relative entropy in terms of asymmetry and coherence}

We conclude the discussion with an application of our quantum speed limit for observables \eqref{eq:ourQSL} in quantum thermodynamics. So far our analysis has been general for any $K$. In the remainder, we will now identify a specific observable $K$ that is of particular thermodynamic relevance. 

Arguably, the central quantity of quantum thermodynamic is the entropy production. Consider a quantum system that is initially prepared in equilibrium and then undergoes a unitary, non-equilibrium process. In this case, the nonequilibrium entropy production is \cite{Deffner2010PRL,Deffner2011PRL,Deffner2013PRE}
\begin{equation}
\label{eq:entropy}
\Sigma=S(\varrho(\tau)||\sigma)\,,
\end{equation}
where $\sigma$ is the thermal Gibbs state corresponding to the initial reference Hamiltonian $H_b$, namely $\sigma=\exp{(-\beta H_b))}/Z$ and $\beta$ is the inverse temperature. 

In Eq.~\eqref{eq:entropy} the entropy production is expressed as the quantum relative entropy \cite{Umegaki1962}, which reads explicitly
\begin{equation}
S(\varrho(\tau)||\sigma)=\tr{\varrho(\tau)\ln(\varrho(\tau))}-\tr{\varrho(\tau)\ln(\sigma)}\,.
\end{equation}
Note that the relative entropy is a measure of distinguishability of quantum states \cite{Umegaki1962}.

Interestingly, Gibbs states are also so-called passive states \cite{Gour2025} that carry neither classical nor quantum correlations \cite{Touil2022JPA}. Thus, in a certain sense, the entropy production $\Sigma$ measures the ``athermality'' of a quantum state $\varrho(t)$ relative to a reference Hamiltonian.

For our present purposes, it is even more interesting that the rate of change of $\Sigma$ was suggested already by Spohn \cite{Spohn1978JMP} as the natural notion of entropy production rate. In particular, we have for unitary dynamics
\begin{equation}
\dot{\Sigma}=-\tr{\dot{\varrho}(t)\ln(\sigma)}\,.
\end{equation}
Thus, we identify $K=-\ln(\sigma)$ as the thermodynamically motivated observable. 

For this case, our quantum speed limit for observables \eqref{eq:ourQSL} becomes
\begin{equation}
\label{eq:QSL_thermo}
v_\mrm{QSL}\leq\frac{1}{2}\,\|[\varrho(t),\ln(\sigma)]\|_1=\frac{\beta}{2}\,\|[\varrho(t),H_b]\|_1\,,
\end{equation}
which is strongly reminiscent of quantum thermodynamic formulations of the Bremerman-Bekenstein bound \cite{Deffner2010PRL,Deffner2020PRR} and thermodynamically relevant versions of the quantum speed limit \cite{Tang2025}. Therefore, Eq.~\eqref{eq:QSL_thermo} can also be interpreted as another version of the \emph{quantum thermodynamic speed limit}. Interestingly, we also observe that in the high-temperature, semiclassical limit $\beta\ll 1$ the quantum speed limit vanishes. Note that the quantum speed limit \eqref{eq:QSL_thermo} for unitary, isolated processes is measured relative to a thermal equilibrium state. Hence, a vanishing speed in the high temperature limit means nothing but that relative to fully mixed state the quantum speed vanishes.

\section{Concluding remarks}

In the present work, we have derived a quantum speed limit for observables in terms of the quantum asymmetry of the quantum state relative to the observable. This quantum speed limit can be determined from weak quantum measurements, provides a lower bound to the amount of information about a parameter conjugate to the observable, and directly related to the coherence encoded in the time-dependent quantum state relative to the eigenbasis of the observable. Interestingly, this novel formulation of the quantum speed limit can also be used to express a quantum thermodynamic speed limit for the nonequilibrium entropy production.

Our results strengthen the intuition that quantum asymmetry and coherence are important quantum resources. In future work, it will be interesting to investigate whether certain measures of general quantum correlation and entanglement can also be leveraged to derive upper bounds on quantum speed. In particular, it  will be interesting to study whether asymmetry, coherence, and entanglement arise in other notions of speed, such as speed of basis rotation \cite{Naseri2024NJP} which are crucial in different protocols of quantum technology.

\acknowledgments{A.B. acknowledges the support from the Fulbright Visiting Scholar Program sponsored by the US Department of State during the completion the present work. AB would also like to thank ICTP Asean Network and PCS IBS for a short research visit grant. The work of H.L.P was funded by the postdoctoral program in BRIN (National Agency for Research and Innovation), Indonesia. This work was partly supported by the {\it Lembaga Pengelolaan Dana Pendidikan} (LPDP) under the scheme of {\it Riset dan Inovasi untuk Indonesia Maju} (RIIM) managed by BRIN with the grant number: 61/II.7/HK/2024. S.D. acknowledges support from the John Templeton Foundation under Grant No. 63626. This work was supported by the U.S. Department of Energy, Office of Basic Energy Sciences, Quantum Information Science program in Chemical Sciences, Geosciences, and Biosciences, under Award No. DE-SC0025997.}

\appendix

\section{Quantum speed limit for qubits \label{Proof of equality in eq. 5}}

In this appendix, we show explicitly that the quantum speed limit \eqref{eq:ourQSL} becomes tight for qubits. To this end, consider the observable $K$ on two-dimensional Hilbert space $\mathbb{C}^2$ with the spectral decomposition, 
\begin{equation}
K=k_+\ket{k_+}\bra{k_+}+k_-\ket{k_-}\bra{k_-},
\label{spectral decomposition of the Hermitian observable}
\end{equation}
where $k_+,k_-\in\mathbb{R}$, and the eigenvectors $\{\ket{k_+},\ket{k_-}\}$ comprises an orthonormal basis of $\mathbb{C}^2$. 

Generally, the Hermitian generator $H(t)$ of the unitary on $\mathbb{C}^2$ can be written as  
\begin{equation}
H(t)=h_+\ket{h_+}\bra{h_+}+h_-\ket{h_-}\bra{h_-}.
\label{general Hermitian operator on 2-dimensional Hilbert space}
\end{equation} 
Here, we have omitted the trivial dependence on $t$. The eigenstates, $\ket{h_{\pm}}$, depend on the polar $\alpha\in[0,\pi]$ and the azimuthal $\beta\in[0,2\pi)$ angles of the Bloch sphere, and $\{h\}=\{h_+,h_-\}$, $h_+,h_-\in\mathbb{R}$, are the corresponding eigenvalues. 

Thus, we can also write
\begin{equation}
\begin{split}
\ket{h_+}=&\cos\left(\alpha/2\right)\,\ket{k_+}+e^{i\beta}\sin\left(\alpha/2\right)\,\ket{k_-}\\
\ket{h_-}=&\sin\left(\alpha/2\right)\,\ket{k_+}-e^{i\beta}\cos\left(\alpha/2\right)\,\ket{k_-}. 
\end{split}
\label{complete set of basis in the x-y plane}
\end{equation}
We can assume, without loss of generality, that $|h_+|>|h_-|$ so that $\|H(t)\|_{\infty}=|h_+|=1$. Then, computing the left-hand side of the inequality in Eq. (\ref{eq:ourQSL}), we obtain 
\begin{widetext}
\begin{eqnarray}
\begin{split}
v_{\rm QSL}=&\max_{\{(h_+,h_-)\},\{(\alpha,\beta)\} }\left\{\frac{|h_+-h_-|}{2}\,|k_+-k_-||\braket{k_-|\varrho|k_+}||\sin(\alpha)\sin(\beta+\phi)|\right\}\\
=&|k_+-k_-||\braket{k_+|\varrho|k_-}|=\|[\varrho,K]\|_1/2.
\end{split}
\label{trace-norm asymmetry vs average noncommutativity for a single qubit}
\end{eqnarray} 
\end{widetext}
Here, $\phi=\arg\braket{k_+|\varrho|k_-}$, the maximum over angular variables $\{(\alpha,\beta)\in[0,\pi]\times[0,2\pi)\}$ is attained for $\alpha=\pi/2$ and $\beta=\pi/2-\phi$, and the maximum over $\{(h_+,h_-)\in\mathbb{R}^2||h_-|\le|h_+|=1\}$ is attained when $h_-=-h_+$. This means that the generator $H(t)$ of the unitary must have eigenvalues $\{-1,1\}$, and the associated pair of orthogonal eigenvectors must lie on the $xy$ plane of the Bloch sphere. The last equality can be checked straightforwardly by using the spectral decomposition of $K$. 

\bibliography{QSL}

@article{Carabba:2022aa,
	author = {Nicoletta Carabba and Niklas H{\"o}rnedal and Adolfo del Campo},
	doi = {https://doi.org/10.22331/q-2022-12-22-884},
	eprint = {2207.05769},
	journal = {Quantum},
	pages = {884},
	title = {Quantum speed limits on operator flows and correlation functions},
	url = {https://arxiv.org/pdf/2207.05769.pdf},
	volume = {6},
	year = {2022}}

@article{H_rnedal_2022,
	author = {H{\"o}rnedal, Niklas and Carabba, Nicoletta and Matsoukas-Roubeas, Apollonas S. and del Campo, Adolfo},
	doi = {10.1038/s42005-022-00985-1},
	issn = {2399-3650},
	journal = {Communications Physics},
	month = aug,
	number = {1},
	publisher = {Springer Science and Business Media LLC},
	title = {Ultimate speed limits to the growth of operator complexity},
	url = {http://dx.doi.org/10.1038/s42005-022-00985-1},
	volume = {5},
	year = {2022}}

@article{Sekiguchi:2024aa,
	author = {Kotaro Sekiguchi and Satoshi Nakajima and Ken Funo and Hiroyasu Tajima},
	month = {10},
	title = {Improvement of Speed Limits: Quantum Effect on the Speed in Open Quantum Systems},
    journal={arXiv preprint arXiv:2410.11604},
	url = {https://arxiv.org/pdf/2410.11604.pdf},
	year = {2024}
}

@article{Budiyono:2021aa,
	author = {Agung Budiyono and Hermawan K. Dipojono},
	doi = {https://doi.org/10.1103/PhysRevA.103.022215},
	eprint = {2106.11436},
	journal = {Physical Review A},
	pages = {022215},
	title = {Quantum uncertainty as classical uncertainty of real-deterministic variables constructed from complex weak values and a global random variable},
	url = {https://arxiv.org/pdf/2106.11436.pdf},
	volume = {103},
	year = {2021}}

@article{Budiyono2023JPAab,
	author = {Agung Budiyono and Bobby E. Gunara and Bagus E. B. Nurhandoko and Hermawan K. Dipojono},
	doi = {https://doi.org/10.1088/1751-8121/acfc04},
	eprint = {2208.03442},
	journal = {J. Phys. A: Math. Theor. 56 435301 (2023)},
	month = {08},
	title = {General quantum correlation from nonreal values of Kirkwood-Dirac quasiprobability over orthonormal product bases},
	url = {https://arxiv.org/pdf/2208.03442.pdf},
	year = {2022}}

@article{Frey2016QINP,
	author = {Frey, Michael R.},
	date = {2016/10/01},
	doi = {10.1007/s11128-016-1405-x},
	id = {Frey2016},
	isbn = {1573-1332},
	journal = {Quantum Inf. Process.},
	number = {10},
	pages = {3919--3950},
	title = {Quantum speed limits---primer, perspectives, and potential future directions},
	url = {https://doi.org/10.1007/s11128-016-1405-x},
	volume = {15},
	year = {2016}}

@article{Deffner2017JPA,
	author = {Deffner, Sebastian and Campbell, Steve},
	doi = {10.1088/1751-8121/aa86c6},
	journal = {J. Phys. A: Math. Theor.},
	month = {oct},
	number = {45},
	pages = {453001},
	publisher = {IOP Publishing},
	title = {Quantum speed limits: from Heisenberg's uncertainty principle to optimal quantum control},
	url = {https://doi.org/10.1088/1751-8121/aa86c6},
	volume = {50},
	year = {2017}}

@article{Naseri2024NJP,
	author = {Naseri, Moein and Macchiavello, Chiara and Bru{\ss}, Dagmar and Horodecki, Pawe{\l} and Streltsov, Alexander},
	doi = {10.1088/1367-2630/ad25a5},
	journal = {New J. Phys.},
	month = {feb},
	number = {2},
	pages = {023052},
	publisher = {IOP Publishing},
	title = {Quantum speed limits for change of basis},
	url = {https://doi.org/10.1088/1367-2630/ad25a5},
	volume = {26},
	year = {2024}}

@book{Gour2025,
	author = {Gour, Gilad},
	doi = {10.1017/9781009560870},
	place = {Cambridge},
	publisher = {Cambridge University Press},
	title = {Quantum Resource Theories},
	year = {2025}}

@inproceedings{Umegaki1962,
	author = {Umegaki, Hisaharu},
	booktitle = {Kodai Mathematical Seminar Reports},
	doi = {10.2996/kmj/1138844604},
	number = {2},
	organization = {Department of Mathematics, Tokyo Institute of Technology},
	pages = {59--85},
	title = {Conditional expectation in an operator algebra, IV (entropy and information)},
	volume = {14},
	year = {1962}}

@article{Mondal2017PRA,
	author = {Mondal, Debasis and Pramanik, Tanumoy and Pati, Arun Kumar},
	doi = {10.1103/PhysRevA.95.010301},
	issue = {1},
	journal = {Phys. Rev. A},
	month = {Jan},
	numpages = {5},
	pages = {010301},
	publisher = {American Physical Society},
	title = {Nonlocal advantage of quantum coherence},
	url = {https://link.aps.org/doi/10.1103/PhysRevA.95.010301},
	volume = {95},
	year = {2017}}

@article{Cheng2015PRA,
	author = {Cheng, Shuming and Hall, Michael J. W.},
	doi = {10.1103/PhysRevA.92.042101},
	issue = {4},
	journal = {Phys. Rev. A},
	month = {Oct},
	numpages = {8},
	pages = {042101},
	publisher = {American Physical Society},
	title = {Complementarity relations for quantum coherence},
	url = {https://link.aps.org/doi/10.1103/PhysRevA.92.042101},
	volume = {92},
	year = {2015}}

@article{Wootters1989AP,
	author = {William K Wootters and Brian D Fields},
	doi = {https://doi.org/10.1016/0003-4916(89)90322-9},
	issn = {0003-4916},
	journal = {Annals of Physics},
	number = {2},
	pages = {363-381},
	title = {Optimal state-determination by mutually unbiased measurements},
	url = {https://www.sciencedirect.com/science/article/pii/0003491689903229},
	volume = {191},
	year = {1989}}

@article{Wootters1986FP,
	author = {Wootters, William K.},
	date = {1986/04/01},
	doi = {10.1007/BF01882696},
	id = {Wootters1986},
	isbn = {1572-9516},
	journal = {Found. Phys.},
	number = {4},
	pages = {391--405},
	title = {Quantum mechanics without probability amplitudes},
	url = {https://doi.org/10.1007/BF01882696},
	volume = {16},
	year = {1986}}

@article{Ivonovic1981JPA,
	author = {I D Ivonovic},
	doi = {10.1088/0305-4470/14/12/019},
	journal = {J. Phys. A: Math. Gen.},
	month = {dec},
	number = {12},
	pages = {3241},
	title = {Geometrical description of quantal state determination},
	url = {https://doi.org/10.1088/0305-4470/14/12/019},
	volume = {14},
	year = {1981}}

@article{Budiyono2024PRA,
	author = {Budiyono, Agung},
	doi = {10.1103/PhysRevA.109.062405},
	issue = {6},
	journal = {Phys. Rev. A},
	month = {Jun},
	numpages = {12},
	pages = {062405},
	publisher = {American Physical Society},
	title = {Sufficient conditions, lower bounds, and trade-off relations for quantumness in Kirkwood-Dirac quasiprobability},
	url = {https://link.aps.org/doi/10.1103/PhysRevA.109.062405},
	volume = {109},
	year = {2024}}

@article{Dirac1945RMP,
	author = {Dirac, P. A. M.},
	doi = {10.1103/RevModPhys.17.195},
	issue = {2-3},
	journal = {Rev. Mod. Phys.},
	month = {Apr},
	numpages = {0},
	pages = {195--199},
	publisher = {American Physical Society},
	title = {On the Analogy Between Classical and Quantum Mechanics},
	url = {https://link.aps.org/doi/10.1103/RevModPhys.17.195},
	volume = {17},
	year = {1945}}

@article{Kirkwood1933PR,
	author = {Kirkwood, John G.},
	doi = {10.1103/PhysRev.44.31},
	issue = {1},
	journal = {Phys. Rev.},
	month = {Jul},
	numpages = {0},
	pages = {31--37},
	publisher = {American Physical Society},
	title = {Quantum Statistics of Almost Classical Assemblies},
	url = {https://link.aps.org/doi/10.1103/PhysRev.44.31},
	volume = {44},
	year = {1933}}

@article{Budiyono2024JPA,
	author = {Budiyono, Agung},
	doi = {10.1088/1751-8121/ad8993},
	journal = {J. Phys. A: Math. Theor.},
	month = {nov},
	number = {46},
	pages = {465303},
	publisher = {IOP Publishing},
	title = {Separation of measurement uncertainty into quantum and classical parts based on Kirkwood--Dirac quasiprobability and generalized entropy},
	url = {https://doi.org/10.1088/1751-8121/ad8993},
	volume = {57},
	year = {2024}}

@article{Hall2023PRA,
	author = {Hall, Michael J. W.},
	doi = {10.1103/PhysRevA.107.062215},
	issue = {6},
	journal = {Phys. Rev. A},
	month = {Jun},
	numpages = {15},
	pages = {062215},
	publisher = {American Physical Society},
	title = {Asymmetry and tighter uncertainty relations for R\'enyi entropies via quantum-classical decompositions of resource measures},
	url = {https://link.aps.org/doi/10.1103/PhysRevA.107.062215},
	volume = {107},
	year = {2023}}

@article{Korzekwa2014PRA,
	author = {Korzekwa, Kamil and Lostaglio, Matteo and Jennings, David and Rudolph, Terry},
	doi = {10.1103/PhysRevA.89.042122},
	issue = {4},
	journal = {Phys. Rev. A},
	month = {Apr},
	numpages = {9},
	pages = {042122},
	publisher = {American Physical Society},
	title = {Quantum and classical entropic uncertainty relations},
	url = {https://link.aps.org/doi/10.1103/PhysRevA.89.042122},
	volume = {89},
	year = {2014}}

@article{Luo2005PRA,
	author = {Luo, Shunlong},
	doi = {10.1103/PhysRevA.72.042110},
	issue = {4},
	journal = {Phys. Rev. A},
	month = {Oct},
	numpages = {3},
	pages = {042110},
	publisher = {American Physical Society},
	title = {Heisenberg uncertainty relation for mixed states},
	url = {https://link.aps.org/doi/10.1103/PhysRevA.72.042110},
	volume = {72},
	year = {2005}}

@article{Braunstein1994PRL,
	author = {Braunstein, Samuel L. and Caves, Carlton M.},
	doi = {10.1103/PhysRevLett.72.3439},
	issue = {22},
	journal = {Phys. Rev. Lett.},
	month = {May},
	numpages = {0},
	pages = {3439--3443},
	publisher = {American Physical Society},
	title = {Statistical distance and the geometry of quantum states},
	url = {https://link.aps.org/doi/10.1103/PhysRevLett.72.3439},
	volume = {72},
	year = {1994}}

@article{Dressel2012PRA,
	author = {Dressel, J. and Jordan, A. N.},
	doi = {10.1103/PhysRevA.85.012107},
	issue = {1},
	journal = {Phys. Rev. A},
	month = {Jan},
	numpages = {13},
	pages = {012107},
	publisher = {American Physical Society},
	title = {Significance of the imaginary part of the weak value},
	url = {https://link.aps.org/doi/10.1103/PhysRevA.85.012107},
	volume = {85},
	year = {2012}}

@article{Aharonov2005PRA,
	author = {Aharonov, Yakir and Botero, Alonso},
	doi = {10.1103/PhysRevA.72.052111},
	issue = {5},
	journal = {Phys. Rev. A},
	month = {Nov},
	numpages = {12},
	pages = {052111},
	publisher = {American Physical Society},
	title = {Quantum averages of weak values},
	url = {https://link.aps.org/doi/10.1103/PhysRevA.72.052111},
	volume = {72},
	year = {2005}}

@article{Kunjwal2019PRA,
	author = {Kunjwal, Ravi and Lostaglio, Matteo and Pusey, Matthew F.},
	doi = {10.1103/PhysRevA.100.042116},
	issue = {4},
	journal = {Phys. Rev. A},
	month = {Oct},
	numpages = {19},
	pages = {042116},
	publisher = {American Physical Society},
	title = {Anomalous weak values and contextuality: Robustness, tightness, and imaginary parts},
	url = {https://link.aps.org/doi/10.1103/PhysRevA.100.042116},
	volume = {100},
	year = {2019}}

@article{Cerezo2021NRP,
	author = {Cerezo, M. and Arrasmith, Andrew and Babbush, Ryan and Benjamin, Simon C. and Endo, Suguru and Fujii, Keisuke and McClean, Jarrod R. and Mitarai, Kosuke and Yuan, Xiao and Cincio, Lukasz and Coles, Patrick J.},
	date = {2021/09/01},
	doi = {10.1038/s42254-021-00348-9},
	id = {Cerezo2021},
	isbn = {2522-5820},
	journal = {Nat. Rev. Phys.},
	number = {9},
	pages = {625--644},
	title = {Variational quantum algorithms},
	url = {https://doi.org/10.1038/s42254-021-00348-9},
	volume = {3},
	year = {2021}}

@article{Chiribella2024PRR,
	author = {Chiribella, Giulio and Simonov, Kyrylo and Zhao, Xuanqiang},
	doi = {10.1103/PhysRevResearch.6.043043},
	issue = {4},
	journal = {Phys. Rev. Res.},
	month = {Oct},
	numpages = {16},
	pages = {043043},
	publisher = {American Physical Society},
	title = {Dimension-independent weak value estimation via controlled SWAP operations},
	url = {https://link.aps.org/doi/10.1103/PhysRevResearch.6.043043},
	volume = {6},
	year = {2024}}

@article{Wagner2024QST,
	author = {Wagner, Rafael and Schwartzman-Nowik, Zohar and Paiva, Ismael L and Te'eni, Amit and Ruiz-Molero, Antonio and Barbosa, Rui Soares and Cohen, Eliahu and Galv{\~a}o, Ernesto F},
	doi = {10.1088/2058-9565/ad124c},
	journal = {Quantum Sci. Technol.},
	month = {jan},
	number = {1},
	pages = {015030},
	publisher = {IOP Publishing},
	title = {Quantum circuits for measuring weak values, Kirkwood--Dirac quasiprobability distributions, and state spectra},
	url = {https://doi.org/10.1088/2058-9565/ad124c},
	volume = {9},
	year = {2024}}

@article{Cohen2018PRA,
	author = {Cohen, Eliahu and Pollak, Eli},
	doi = {10.1103/PhysRevA.98.042112},
	issue = {4},
	journal = {Phys. Rev. A},
	month = {Oct},
	numpages = {8},
	pages = {042112},
	publisher = {American Physical Society},
	title = {Determination of weak values of quantum operators using only strong measurements},
	url = {https://link.aps.org/doi/10.1103/PhysRevA.98.042112},
	volume = {98},
	year = {2018}}

@article{Vallone2016PRL,
	author = {Vallone, Giuseppe and Dequal, Daniele},
	doi = {10.1103/PhysRevLett.116.040502},
	issue = {4},
	journal = {Phys. Rev. Lett.},
	month = {Jan},
	numpages = {5},
	pages = {040502},
	publisher = {American Physical Society},
	title = {Strong Measurements Give a Better Direct Measurement of the Quantum Wave Function},
	url = {https://link.aps.org/doi/10.1103/PhysRevLett.116.040502},
	volume = {116},
	year = {2016}}

@article{Johansen2007PLA,
	author = {Lars M. Johansen},
	doi = {https://doi.org/10.1016/j.physleta.2007.02.039},
	issn = {0375-9601},
	journal = {Phys. Lett. A},
	number = {4},
	pages = {374-376},
	title = {Reconstructing weak values without weak measurements},
	url = {https://www.sciencedirect.com/science/article/pii/S0375960107002654},
	volume = {366},
	year = {2007}}

@article{Jozsa2007PRA,
	author = {Jozsa, Richard},
	doi = {10.1103/PhysRevA.76.044103},
	issue = {4},
	journal = {Phys. Rev. A},
	month = {Oct},
	numpages = {3},
	pages = {044103},
	publisher = {American Physical Society},
	title = {Complex weak values in quantum measurement},
	url = {https://link.aps.org/doi/10.1103/PhysRevA.76.044103},
	volume = {76},
	year = {2007}}

@article{Lundeen2005PLA,
	author = {J.S. Lundeen and K.J. Resch},
	doi = {https://doi.org/10.1016/j.physleta.2004.11.037},
	issn = {0375-9601},
	journal = {Phys. Lett. A},
	number = {5},
	pages = {337-344},
	title = {Practical measurement of joint weak values and their connection to the annihilation operator},
	url = {https://www.sciencedirect.com/science/article/pii/S0375960104016342},
	volume = {334},
	year = {2005}}

@article{Wiseman2002PRA,
	author = {Wiseman, H. M.},
	doi = {10.1103/PhysRevA.65.032111},
	issue = {3},
	journal = {Phys. Rev. A},
	month = {Feb},
	numpages = {6},
	pages = {032111},
	publisher = {American Physical Society},
	title = {Weak values, quantum trajectories, and the cavity-QED experiment on wave-particle correlation},
	url = {https://link.aps.org/doi/10.1103/PhysRevA.65.032111},
	volume = {65},
	year = {2002}}

@article{Aharonov1988PRL,
	author = {Aharonov, Yakir and Albert, David Z. and Vaidman, Lev},
	doi = {10.1103/PhysRevLett.60.1351},
	issue = {14},
	journal = {Phys. Rev. Lett.},
	month = {Apr},
	numpages = {0},
	pages = {1351--1354},
	publisher = {American Physical Society},
	title = {How the result of a measurement of a component of the spin of a spin-1/2 particle can turn out to be 100},
	url = {https://link.aps.org/doi/10.1103/PhysRevLett.60.1351},
	volume = {60},
	year = {1988}}

@article{Wigner1963PNAS,
	author = {Wigner, Eugene P and Yanase, Mutsuo M},
	doi = {10.1073/pnas.49.6.910},
	journal = {PNAS},
	number = {6},
	pages = {910--918},
	title = {Information contents of distributions},
	volume = {49},
	year = {1963}}

@article{Mondal2016PLA,
	author = {Debasis Mondal and Chandan Datta and Sk Sazim},
	doi = {https://doi.org/10.1016/j.physleta.2015.12.015},
	issn = {0375-9601},
	journal = {Phys. Lett. A},
	number = {5},
	pages = {689-695},
	title = {Quantum coherence sets the quantum speed limit for mixed states},
	url = {https://www.sciencedirect.com/science/article/pii/S0375960115010518},
	volume = {380},
	year = {2016}}

@article{Marvian2016PRA,
	author = {Marvian, Iman and Spekkens, Robert W. and Zanardi, Paolo},
	doi = {10.1103/PhysRevA.93.052331},
	issue = {5},
	journal = {Phys. Rev. A},
	month = {May},
	numpages = {12},
	pages = {052331},
	publisher = {American Physical Society},
	title = {Quantum speed limits, coherence, and asymmetry},
	url = {https://link.aps.org/doi/10.1103/PhysRevA.93.052331},
	volume = {93},
	year = {2016}}

@article{Allahverdyan2004EPL,
	author = {A. E. Allahverdyan and R. Balian and Th. M. Nieuwenhuizen},
	doi = {10.1209/epl/i2004-10101-2},
	journal = {EPL (Europhys. Lett.)},
	month = {aug},
	number = {4},
	pages = {565},
	title = {Maximal work extraction from finite quantum systems},
	url = {https://doi.org/10.1209/epl/i2004-10101-2},
	volume = {67},
	year = {2004}}

@article{Budiyono2025PRA,
	author = {Budiyono, Agung},
	doi = {10.1103/PhysRevA.111.012216},
	issue = {1},
	journal = {Phys. Rev. A},
	month = {Jan},
	numpages = {13},
	pages = {012216},
	publisher = {American Physical Society},
	title = {Quantum entanglement as an extremal Kirkwood-Dirac nonreality},
	url = {https://link.aps.org/doi/10.1103/PhysRevA.111.012216},
	volume = {111},
	year = {2025}}

@article{Horodecki2009RMP,
	author = {Horodecki, Ryszard and Horodecki, Pawe\l{} and Horodecki, Micha\l{} and Horodecki, Karol},
	doi = {10.1103/RevModPhys.81.865},
	issue = {2},
	journal = {Rev. Mod. Phys.},
	month = {Jun},
	numpages = {0},
	pages = {865--942},
	publisher = {American Physical Society},
	title = {Quantum entanglement},
	url = {https://link.aps.org/doi/10.1103/RevModPhys.81.865},
	volume = {81},
	year = {2009}}

@article{Adesso2016JPA,
	author = {Adesso, Gerardo and Bromley, Thomas R and Cianciaruso, Marco},
	doi = {10.1088/1751-8113/49/47/473001},
	journal = {J. Phys. A: Math. Theor.},
	month = {nov},
	number = {47},
	pages = {473001},
	publisher = {IOP Publishing},
	title = {Measures and applications of quantum correlations},
	url = {https://doi.org/10.1088/1751-8113/49/47/473001},
	volume = {49},
	year = {2016}}

@article{Budiyono2023PRA,
	author = {Budiyono, Agung and Dipojono, Hermawan K.},
	doi = {10.1103/PhysRevA.107.022408},
	issue = {2},
	journal = {Phys. Rev. A},
	month = {Feb},
	numpages = {9},
	pages = {022408},
	publisher = {American Physical Society},
	title = {Quantifying quantum coherence via Kirkwood-Dirac quasiprobability},
	url = {https://link.aps.org/doi/10.1103/PhysRevA.107.022408},
	volume = {107},
	year = {2023}}

@article{Girolami2014PRL,
	author = {Girolami, Davide},
	doi = {10.1103/PhysRevLett.113.170401},
	issue = {17},
	journal = {Phys. Rev. Lett.},
	month = {Oct},
	numpages = {5},
	pages = {170401},
	publisher = {American Physical Society},
	title = {Observable Measure of Quantum Coherence in Finite Dimensional Systems},
	url = {https://link.aps.org/doi/10.1103/PhysRevLett.113.170401},
	volume = {113},
	year = {2014}}

@article{Baumgratz2014PRL,
	author = {Baumgratz, T. and Cramer, M. and Plenio, M. B.},
	doi = {10.1103/PhysRevLett.113.140401},
	issue = {14},
	journal = {Phys. Rev. Lett.},
	month = {Sep},
	numpages = {5},
	pages = {140401},
	publisher = {American Physical Society},
	title = {Quantifying Coherence},
	url = {https://link.aps.org/doi/10.1103/PhysRevLett.113.140401},
	volume = {113},
	year = {2014}}

@article{Streltsov2017RMP,
	author = {Streltsov, Alexander and Adesso, Gerardo and Plenio, Martin B.},
	doi = {10.1103/RevModPhys.89.041003},
	issue = {4},
	journal = {Rev. Mod. Phys.},
	month = {Oct},
	numpages = {34},
	pages = {041003},
	publisher = {American Physical Society},
	title = {Colloquium: Quantum coherence as a resource},
	url = {https://link.aps.org/doi/10.1103/RevModPhys.89.041003},
	volume = {89},
	year = {2017}}

@article{Budiyono2023JPA,
	author = {Budiyono, Agung and Agusta, Mohammad K and Nurhandoko, Bagus E B and Dipojono, Hermawan K},
	doi = {10.1088/1751-8121/acd091},
	journal = {J. Phys. A: Math. Theor.},
	month = {may},
	number = {23},
	pages = {235304},
	publisher = {IOP Publishing},
	title = {Quantum coherence as asymmetry from complex weak values},
	url = {https://doi.org/10.1088/1751-8121/acd091},
	volume = {56},
	year = {2023}}

@article{Budiyono2023PRA_2,
	author = {Budiyono, Agung},
	doi = {10.1103/PhysRevA.108.012431},
	issue = {1},
	journal = {Phys. Rev. A},
	month = {Jul},
	numpages = {12},
	pages = {012431},
	publisher = {American Physical Society},
	title = {Operational interpretation and estimation of quantum trace-norm asymmetry based on weak-value measurement and some bounds},
	url = {https://link.aps.org/doi/10.1103/PhysRevA.108.012431},
	volume = {108},
	year = {2023}}

@article{Marvian2016PRA_2,
	author = {Marvian, Iman and Spekkens, Robert W.},
	doi = {10.1103/PhysRevA.94.052324},
	issue = {5},
	journal = {Phys. Rev. A},
	month = {Nov},
	numpages = {23},
	pages = {052324},
	publisher = {American Physical Society},
	title = {How to quantify coherence: Distinguishing speakable and unspeakable notions},
	url = {https://link.aps.org/doi/10.1103/PhysRevA.94.052324},
	volume = {94},
	year = {2016}}

@article{Vaccaro2008PRA,
	author = {Vaccaro, J. A. and Anselmi, F. and Wiseman, H. M. and Jacobs, K.},
	doi = {10.1103/PhysRevA.77.032114},
	issue = {3},
	journal = {Phys. Rev. A},
	month = {Mar},
	numpages = {12},
	pages = {032114},
	publisher = {American Physical Society},
	title = {Tradeoff between extractable mechanical work, accessible entanglement, and ability to act as a reference system, under arbitrary superselection rules},
	url = {https://link.aps.org/doi/10.1103/PhysRevA.77.032114},
	volume = {77},
	year = {2008}}

@article{Gour2008NJP,
	author = {Gour, Gilad and Spekkens, Robert W},
	doi = {10.1088/1367-2630/10/3/033023},
	journal = {New J. Phys.},
	month = {mar},
	number = {3},
	pages = {033023},
	title = {The resource theory of quantum reference frames: manipulations and monotones},
	url = {https://doi.org/10.1088/1367-2630/10/3/033023},
	volume = {10},
	year = {2008}}

@article{Shrimali2024PRA,
	author = {Shrimali, Divyansh and Panda, Biswaranjan and Pati, Arun Kumar},
	doi = {10.1103/PhysRevA.110.022425},
	issue = {2},
	journal = {Phys. Rev. A},
	month = {Aug},
	numpages = {14},
	pages = {022425},
	publisher = {American Physical Society},
	title = {Stronger speed limit for observables: Tighter bound for the capacity of entanglement, the modular Hamiltonian, and the charging of a quantum battery},
	url = {https://link.aps.org/doi/10.1103/PhysRevA.110.022425},
	volume = {110},
	year = {2024}}

@article{Pintos2022PRX,
	author = {Garc\'{\i}a-Pintos, Luis Pedro and Nicholson, Schuyler B. and Green, Jason R. and del Campo, Adolfo and Gorshkov, Alexey V.},
	doi = {10.1103/PhysRevX.12.011038},
	issue = {1},
	journal = {Phys. Rev. X},
	month = {Feb},
	numpages = {22},
	pages = {011038},
	publisher = {American Physical Society},
	title = {Unifying Quantum and Classical Speed Limits on Observables},
	url = {https://link.aps.org/doi/10.1103/PhysRevX.12.011038},
	volume = {12},
	year = {2022}}

@article{Mohan2022PRA,
	author = {Mohan, Brij and Pati, Arun Kumar},
	doi = {10.1103/PhysRevA.106.042436},
	issue = {4},
	journal = {Phys. Rev. A},
	month = {Oct},
	numpages = {12},
	pages = {042436},
	publisher = {American Physical Society},
	title = {Quantum speed limits for observables},
	url = {https://link.aps.org/doi/10.1103/PhysRevA.106.042436},
	volume = {106},
	year = {2022}}

@article{Deffner2010PRL,
	author = {Deffner, Sebastian and Lutz, Eric},
	doi = {10.1103/PhysRevLett.105.170402},
	issue = {17},
	journal = {Phys. Rev. Lett.},
	month = {Oct},
	numpages = {4},
	pages = {170402},
	publisher = {American Physical Society},
	title = {Generalized Clausius Inequality for Nonequilibrium Quantum Processes},
	url = {https://link.aps.org/doi/10.1103/PhysRevLett.105.170402},
	volume = {105},
	year = {2010}}

@article{Campbell2018QST,
	author = {Campbell, Steve and Genoni, Marco G and Deffner, Sebastian},
	doi = {10.1088/2058-9565/aaa641},
	journal = {Quantum Sci. Technol.},
	month = {feb},
	number = {2},
	pages = {025002},
	publisher = {IOP Publishing},
	title = {Precision thermometry and the quantum speed limit},
	url = {https://doi.org/10.1088/2058-9565/aaa641},
	volume = {3},
	year = {2018}}

@article{Deffner2013PRE,
	author = {Deffner, Sebastian and Lutz, Eric},
	doi = {10.1103/PhysRevE.87.022143},
	issue = {2},
	journal = {Phys. Rev. E},
	month = {Feb},
	numpages = {7},
	pages = {022143},
	publisher = {American Physical Society},
	title = {Thermodynamic length for far-from-equilibrium quantum systems},
	url = {https://link.aps.org/doi/10.1103/PhysRevE.87.022143},
	volume = {87},
	year = {2013}}

@article{Deffner2011PRL,
	author = {Deffner, Sebastian and Lutz, Eric},
	doi = {10.1103/PhysRevLett.107.140404},
	issue = {14},
	journal = {Phys. Rev. Lett.},
	month = {Sep},
	numpages = {5},
	pages = {140404},
	publisher = {American Physical Society},
	title = {Nonequilibrium Entropy Production for Open Quantum Systems},
	url = {https://link.aps.org/doi/10.1103/PhysRevLett.107.140404},
	volume = {107},
	year = {2011}}

@article{Touil2022JPA,
	author = {Touil, Akram and {\c C}akmak, Barı{\c s} and Deffner, Sebastian},
	doi = {10.1088/1751-8121/ac3eba},
	journal = {J. Phys. A: Math. Theor.},
	month = {dec},
	number = {2},
	pages = {025301},
	publisher = {IOP Publishing},
    title={Ergotropy from quantum and classical correlations},
	url = {https://doi.org/10.1088/1751-8121/ac3eba},
	volume = {55},
	year = {2021}}

@article{Giovannetti2011NPhot,
	author = {Giovannetti, Vittorio and Lloyd, Seth and Maccone, Lorenzo},
	date = {2011/04/01},
	doi = {10.1038/nphoton.2011.35},
	id = {Giovannetti2011},
	isbn = {1749-4893},
	journal = {Nature Photon.},
	number = {4},
	pages = {222--229},
	title = {Advances in quantum metrology},
	url = {https://doi.org/10.1038/nphoton.2011.35},
	volume = {5},
	year = {2011}}

@article{Mukhopadhyay2018PRE,
	author = {Mukhopadhyay, Chiranjib and Misra, Avijit and Bhattacharya, Samyadeb and Pati, Arun Kumar},
	doi = {10.1103/PhysRevE.97.062116},
	issue = {6},
	journal = {Phys. Rev. E},
	month = {Jun},
	numpages = {8},
	pages = {062116},
	publisher = {American Physical Society},
	title = {Quantum speed limit constraints on a nanoscale autonomous refrigerator},
	url = {https://link.aps.org/doi/10.1103/PhysRevE.97.062116},
	volume = {97},
	year = {2018}}

@article{Funo2019NJP,
	author = {Funo, Ken and Shiraishi, Naoto and Saito, Keiji},
	doi = {10.1088/1367-2630/aaf9f5},
	journal = {New J. Phys.},
	month = {jan},
	number = {1},
	pages = {013006},
	publisher = {IOP Publishing},
	title = {Speed limit for open quantum systems},
	url = {https://doi.org/10.1088/1367-2630/aaf9f5},
	volume = {21},
	year = {2019}}

@article{Campbell2017PRL,
	author = {Campbell, Steve and Deffner, Sebastian},
	doi = {10.1103/PhysRevLett.118.100601},
	issue = {10},
	journal = {Phys. Rev. Lett.},
	month = {Mar},
	numpages = {7},
	pages = {100601},
	publisher = {American Physical Society},
	title = {Trade-Off Between Speed and Cost in Shortcuts to Adiabaticity},
	url = {https://link.aps.org/doi/10.1103/PhysRevLett.118.100601},
	volume = {118},
	year = {2017}}

@article{Canvea2009PRL,
	author = {Caneva, T. and Murphy, M. and Calarco, T. and Fazio, R. and Montangero, S. and Giovannetti, V. and Santoro, G. E.},
	doi = {10.1103/PhysRevLett.103.240501},
	issue = {24},
	journal = {Phys. Rev. Lett.},
	month = {Dec},
	numpages = {4},
	pages = {240501},
	publisher = {American Physical Society},
	title = {Optimal Control at the Quantum Speed Limit},
	url = {https://link.aps.org/doi/10.1103/PhysRevLett.103.240501},
	volume = {103},
	year = {2009}}

@article{Ashhab2012PRA,
	author = {Ashhab, S. and de Groot, P. C. and Nori, Franco},
	doi = {10.1103/PhysRevA.85.052327},
	issue = {5},
	journal = {Phys. Rev. A},
	month = {May},
	numpages = {7},
	pages = {052327},
	publisher = {American Physical Society},
	title = {Speed limits for quantum gates in multiqubit systems},
	url = {https://link.aps.org/doi/10.1103/PhysRevA.85.052327},
	volume = {85},
	year = {2012}}

@article{Lloyd2000Nature,
	author = {Lloyd, Seth},
	date = {2000/08/01},
	doi = {10.1038/35023282},
	id = {Lloyd2000},
	isbn = {1476-4687},
	journal = {Nature},
	number = {6799},
	pages = {1047--1054},
	title = {Ultimate physical limits to computation},
	url = {https://doi.org/10.1038/35023282},
	volume = {406},
	year = {2000}}

@article{Mohan2022NJP,
	author = {Mohan, Brij and Das, Siddhartha and Pati, Arun Kumar},
	doi = {10.1088/1367-2630/ac753c},
	journal = {New J. Phys.},
	month = {jun},
	number = {6},
	pages = {065003},
	publisher = {IOP Publishing},
	title = {Quantum speed limits for information and coherence},
	url = {https://doi.org/10.1088/1367-2630/ac753c},
	volume = {24},
	year = {2022}}

@article{Hamazaki2022PRXQ,
	author = {Hamazaki, Ryusuke},
	doi = {10.1103/PRXQuantum.3.020319},
	issue = {2},
	journal = {PRX Quantum},
	month = {Apr},
	numpages = {37},
	pages = {020319},
	publisher = {American Physical Society},
	title = {Speed Limits for Macroscopic Transitions},
	url = {https://link.aps.org/doi/10.1103/PRXQuantum.3.020319},
	volume = {3},
	year = {2022}}

@article{Baumgartner2011,
	author = {Bernhard Baumgartner},
	doi = {10.48550/arXiv.1106.6189},
	journal = {arXiv preprint arXiv:1106.6189},
	title = {An inequality for the trace of matrix products, using absolute values},
	year = {2011}}

@article{Spohn1978JMP,
	author = {Spohn, Herbert},
	doi = {10.1063/1.523789},
	issn = {0022-2488},
	journal = {J. Math. Phys.},
	month = {05},
	number = {5},
	pages = {1227-1230},
	title = {Entropy production for quantum dynamical semigroups},
	volume = {19},
	year = {1978}}

@article{OConnor2021PRA,
	author = {O'Connor, Eoin and Guarnieri, Giacomo and Campbell, Steve},
	doi = {10.1103/PhysRevA.103.022210},
	issue = {2},
	journal = {Phys. Rev. A},
	month = {Feb},
	numpages = {8},
	pages = {022210},
	publisher = {American Physical Society},
	title = {Action quantum speed limits},
	url = {https://link.aps.org/doi/10.1103/PhysRevA.103.022210},
	volume = {103},
	year = {2021}}

@article{Campaioli2018PRL,
	author = {Campaioli, Francesco and Pollock, Felix A. and Binder, Felix C. and Modi, Kavan},
	doi = {10.1103/PhysRevLett.120.060409},
	issue = {6},
	journal = {Phys. Rev. Lett.},
	month = {Feb},
	numpages = {5},
	pages = {060409},
	publisher = {American Physical Society},
	title = {Tightening Quantum Speed Limits for Almost All States},
	url = {https://link.aps.org/doi/10.1103/PhysRevLett.120.060409},
	volume = {120},
	year = {2018}}

@article{Pires2016PRX,
	author = {Pires, Diego Paiva and Cianciaruso, Marco and C\'eleri, Lucas C. and Adesso, Gerardo and Soares-Pinto, Diogo O.},
	doi = {10.1103/PhysRevX.6.021031},
	issue = {2},
	journal = {Phys. Rev. X},
	month = {Jun},
	numpages = {19},
	pages = {021031},
	publisher = {American Physical Society},
	title = {Generalized Geometric Quantum Speed Limits},
	url = {https://link.aps.org/doi/10.1103/PhysRevX.6.021031},
	volume = {6},
	year = {2016}}

@article{Deffner2013PRL,
	author = {Deffner, Sebastian and Lutz, Eric},
	doi = {10.1103/PhysRevLett.111.010402},
	issue = {1},
	journal = {Phys. Rev. Lett.},
	month = {Jul},
	numpages = {5},
	pages = {010402},
	publisher = {American Physical Society},
	title = {Quantum Speed Limit for Non-Markovian Dynamics},
	url = {https://link.aps.org/doi/10.1103/PhysRevLett.111.010402},
	volume = {111},
	year = {2013}}

@article{Campo2013PRL,
	author = {del Campo, A. and Egusquiza, I. L. and Plenio, M. B. and Huelga, S. F.},
	doi = {10.1103/PhysRevLett.110.050403},
	issue = {5},
	journal = {Phys. Rev. Lett.},
	month = {Jan},
	numpages = {5},
	pages = {050403},
	publisher = {American Physical Society},
	title = {Quantum Speed Limits in Open System Dynamics},
	url = {https://link.aps.org/doi/10.1103/PhysRevLett.110.050403},
	volume = {110},
	year = {2013}}

@article{Taddei2013PRL,
	author = {Taddei, M. M. and Escher, B. M. and Davidovich, L. and de Matos Filho, R. L.},
	doi = {10.1103/PhysRevLett.110.050402},
	issue = {5},
	journal = {Phys. Rev. Lett.},
	month = {Jan},
	numpages = {5},
	pages = {050402},
	publisher = {American Physical Society},
	title = {Quantum Speed Limit for Physical Processes},
	url = {https://link.aps.org/doi/10.1103/PhysRevLett.110.050402},
	volume = {110},
	year = {2013}}

@article{Levitin2009PRL,
	author = {Levitin, Lev B. and Toffoli, Tommaso},
	doi = {10.1103/PhysRevLett.103.160502},
	issue = {16},
	journal = {Phys. Rev. Lett.},
	month = {Oct},
	numpages = {4},
	pages = {160502},
	publisher = {American Physical Society},
	title = {Fundamental Limit on the Rate of Quantum Dynamics: The Unified Bound Is Tight},
	url = {https://link.aps.org/doi/10.1103/PhysRevLett.103.160502},
	volume = {103},
	year = {2009}}

@article{Jones2010PRA,
	author = {Jones, Philip J. and Kok, Pieter},
	doi = {10.1103/PhysRevA.82.022107},
	issue = {2},
	journal = {Phys. Rev. A},
	month = {Aug},
	numpages = {7},
	pages = {022107},
	publisher = {American Physical Society},
	title = {Geometric derivation of the quantum speed limit},
	url = {https://link.aps.org/doi/10.1103/PhysRevA.82.022107},
	volume = {82},
	year = {2010}}

@article{Kupferman2008PRA,
	author = {Kupferman, Judy and Reznik, Benni},
	doi = {10.1103/PhysRevA.78.042305},
	issue = {4},
	journal = {Phys. Rev. A},
	month = {Oct},
	numpages = {7},
	pages = {042305},
	publisher = {American Physical Society},
	title = {Entanglement and the speed of evolution in mixed states},
	url = {https://link.aps.org/doi/10.1103/PhysRevA.78.042305},
	volume = {78},
	year = {2008}}

@article{Margolus1998PD,
	author = {Norman Margolus and Lev B. Levitin},
	doi = {https://doi.org/10.1016/S0167-2789(98)00054-2},
	issn = {0167-2789},
	journal = {Physica D},
	note = {Proceedings of the Fourth Workshop on Physics and Consumption},
	number = {1},
	pages = {188-195},
	title = {The maximum speed of dynamical evolution},
	url = {https://www.sciencedirect.com/science/article/pii/S0167278998000542},
	volume = {120},
	year = {1998}}

@article{Uffink1993AJP,
	author = {Uffink, Jos},
	doi = {10.1119/1.17368},
	issn = {0002-9505},
	journal = {Am. J. Phys.},
	month = {10},
	number = {10},
	pages = {935-936},
	title = {The rate of evolution of a quantum state},
	url = {https://doi.org/10.1119/1.17368},
	volume = {61},
	year = {1993}}

@article{Campaioli2022NJP,
	author = {Campaioli, Francesco and Yu, Chang-shui and Pollock, Felix A and Modi, Kavan},
	doi = {10.1088/1367-2630/ac7346},
	journal = {New J. Phys.},
	month = {jun},
	number = {6},
	pages = {065001},
	publisher = {IOP Publishing},
	title = {Resource speed limits: maximal rate of resource variation},
	url = {https://doi.org/10.1088/1367-2630/ac7346},
	volume = {24},
	year = {2022}}

@article{Allan2021Quantum,
	author = {Allan, Dan and H{\"{o}}rnedal, Niklas and Andersson, Ole},
	doi = {10.22331/q-2021-05-27-462},
	issn = {2521-327X},
	journal = {{Quantum}},
	month = may,
	pages = {462},
	publisher = {{Verein zur F{\"{o}}rderung des Open Access Publizierens in den Quantenwissenschaften}},
	title = {Time-optimal quantum transformations with bounded bandwidth},
	url = {https://doi.org/10.22331/q-2021-05-27-462},
	volume = {5},
	year = {2021}}

@article{Pratapsi2025QST,
	author = {Silva Pratapsi, Sagar and Deffner, Sebastian and Gherardini, Stefano},
	doi = {10.1088/2058-9565/add55d},
	journal = {Quantum Sci. Technol.},
	month = {may},
	number = {3},
	pages = {035019},
	publisher = {IOP Publishing},
	title = {Quantum speed limit for Kirkwood--Dirac quasiprobabilities},
	url = {https://doi.org/10.1088/2058-9565/add55d},
	volume = {10},
	year = {2025}}

@article{Vaidman1992AJP,
	author = {Vaidman, Lev},
	doi = {10.1119/1.16940},
	issn = {0002-9505},
	journal = {Am. J. Phys.},
	month = {02},
	number = {2},
	pages = {182-183},
	title = {Minimum time for the evolution to an orthogonal quantum state},
	url = {https://doi.org/10.1119/1.16940},
	volume = {60},
	year = {1992}}

@article{Fleming1973,
	author = {Fleming, G. N.},
	date = {1973/07/01},
	doi = {10.1007/BF02819419},
	id = {Fleming1973},
	isbn = {1826-9869},
	journal = {Nuov. Cim. A},
	number = {2},
	pages = {232--240},
	title = {A unitarity bound on the evolution of nonstationary states},
	url = {https://doi.org/10.1007/BF02819419},
	volume = {16},
	year = {1973}}

@article{Bhattacharyya1983JPA,
	author = {K Bhattacharyya},
	doi = {10.1088/0305-4470/16/13/021},
	journal = {J. Phys. A: Math. Gen.},
	month = {sep},
	number = {13},
	pages = {2993},
	title = {Quantum decay and the Mandelstam-Tamm-energy inequality},
	url = {https://doi.org/10.1088/0305-4470/16/13/021},
	volume = {16},
	year = {1983}}

@article{Anandan1990PRL,
	author = {Anandan, J. and Aharonov, Y.},
	doi = {10.1103/PhysRevLett.65.1697},
	issue = {14},
	journal = {Phys. Rev. Lett.},
	month = {Oct},
	numpages = {0},
	pages = {1697--1700},
	publisher = {American Physical Society},
	title = {Geometry of quantum evolution},
	url = {https://link.aps.org/doi/10.1103/PhysRevLett.65.1697},
	volume = {65},
	year = {1990}}

@article{Pati1991PLA,
	author = {Arun Kumar Pati},
	doi = {https://doi.org/10.1016/0375-9601(91)90255-7},
	issn = {0375-9601},
	journal = {Phys. Lett. A},
	number = {3},
	pages = {105-112},
	title = {Relation between ``phases'' and ``distance'' in quantum evolution},
	url = {https://www.sciencedirect.com/science/article/pii/0375960191902557},
	volume = {159},
	year = {1991}}

@article{Uhlmann1992PLA,
	author = {Armin Uhlmann},
	doi = {https://doi.org/10.1016/0375-9601(92)90555-Z},
	issn = {0375-9601},
	journal = {Phys. Lett. A},
	number = {4},
	pages = {329-331},
	title = {An energy dispersion estimate},
	url = {https://www.sciencedirect.com/science/article/pii/037596019290555Z},
	volume = {161},
	year = {1992}}

@article{Mandelstam1945,
	author = {Mandelstam, L. and Tamm, I.},
	doi = {10.1007/978-3-642-74626-0_8},
	journal = {J. Phys.},
	pages = {249},
	title = {The uncertainty relation between energy and time in nonrelativistic quantum mechanics},
	url = {https://link.springer.com/chapter/10.1007/978-3-642-74626-0_8},
	volume = {9},
	year = {1945}}

@article{Deffner2025QST,
	author = {Deffner, Sebastian},
	doi = {10.1088/2058-9565/adac05},
	journal = {Quantum Sci. Technol.},
	month = {jan},
	number = {2},
	pages = {025009},
	publisher = {IOP Publishing},
	title = {Towards enhanced precision in thermometry with nonlinear qubits},
	url = {https://doi.org/10.1088/2058-9565/adac05},
	volume = {10},
	year = {2025}}

@article{Deffner2022EPL,
	author = {Deffner, Sebastian},
	doi = {10.1209/0295-5075/ac9fed},
	journal = {EPL (Europhys. Lett.)},
	month = {nov},
	number = {4},
	pages = {48001},
	publisher = {EDP Sciences, IOP Publishing and Societ{\`a} Italiana di Fisica},
	title = {Nonlinear speed-ups in ultracold quantum gases},
	url = {https://doi.org/10.1209/0295-5075/ac9fed},
	volume = {140},
	year = {2022}}

@article{Aifer2022NJP,
	author = {Aifer, Maxwell and Deffner, Sebastian},
	doi = {10.1088/1367-2630/ac6821},
	journal = {New J. Phys.},
	month = {may},
	number = {5},
	pages = {055002},
	publisher = {IOP Publishing},
	title = {From quantum speed limits to energy-efficient quantum gates},
	url = {https://doi.org/10.1088/1367-2630/ac6821},
	volume = {24},
	year = {2022}}

@article{Poggi2021PRXQ,
	author = {Poggi, Pablo M. and Campbell, Steve and Deffner, Sebastian},
	doi = {10.1103/PRXQuantum.2.040349},
	issue = {4},
	journal = {PRX Quantum},
	month = {Dec},
	numpages = {11},
	pages = {040349},
	publisher = {American Physical Society},
	title = {Diverging Quantum Speed Limits: A Herald of Classicality},
	url = {https://link.aps.org/doi/10.1103/PRXQuantum.2.040349},
	volume = {2},
	year = {2021}}

@article{Tang2025,
	author = {Tang, Shou-I and Doucet, Emery and Touil, Akram and Deffner, Sebastian and Sone, Akira},
	doi = {arXiv.2511.08858},
	journal = {arXiv preprint arXiv:2511.08858},
	title = {Information Processing in Quantum Thermodynamic Systems: an Autonomous Hamiltonian Approach},
	year = {2025}}

@article{Deffner2020PRR,
	author = {Deffner, Sebastian},
	doi = {10.1103/PhysRevResearch.2.013161},
	issue = {1},
	journal = {Phys. Rev. Res.},
	month = {Feb},
	numpages = {7},
	pages = {013161},
	publisher = {American Physical Society},
	title = {Quantum speed limits and the maximal rate of information production},
	url = {https://link.aps.org/doi/10.1103/PhysRevResearch.2.013161},
	volume = {2},
	year = {2020}}

\end{document}